\documentclass[a4paper,11pt]{article}
\usepackage{jheppub} % for details on the use of the package, please see the JINST-author-manual
\usepackage{lineno}
\RequirePackage{fix-cm}
\RequirePackage{graphicx}
\RequirePackage{amsmath}
\RequirePackage{booktabs}
\RequirePackage{upgreek}
\RequirePackage{xspace}
\RequirePackage{orcidlink}

\newcommand{\SND}{SND@LHC\xspace}
\newcommand{\numu}{\nu_\mu}
\newcommand{\numubar}{\bar{\nu}_\mu}
\newcommand{\fbinv}{\,\text{fb}^{-1}}
\newcommand{\abinv}{\,\text{ab}^{-1}}
\newcommand{\sqrts}{\sqrt{s}}

\newcommand{\GeV}{\,\text{GeV}}
\newcommand{\TeV}{\,\text{TeV}}
\newcommand{\muhat}{\hat{\mu}}

\arxivnumber{1234.56789} % if you have one

\title{Measurement of the muon neutrino charged-current cross section with \SND}

\collaboration{The \SND{} Collaboration}
\author[9]{D.~Abbaneo\orcidlink{0000-0001-9416-1742}}
\author[43]{S.~Ahmad\orcidlink{0000-0001-8236-6134}}
\author[1,2]{R.~Albanese\orcidlink{0000-0003-4586-8068}}
\author[1]{A.~Alexandrov\orcidlink{0000-0002-1813-1485}}
\author[1,2]{F.~Alicante\orcidlink{0009-0003-3240-830X}}
\author[1,2]{F.~Aloschi\orcidlink{0000-0002-2501-7525}}
\author[6]{K.~Androsov\orcidlink{0000-0003-2694-6542}}
\author[1,2]{L.G.~Arellano\orcidlink{0000-0002-1093-1824}}
\author[39]{C.~Asawatangtrakuldee\orcidlink{0000-0003-2234-7219}}
\author[28,33]{M.A.~Ayala~Torres\orcidlink{0000-0002-4296-9464}}
% ...

\affiliation[1]{Sezione INFN di Napoli, Napoli, 80126, Italy}
\affiliation[2]{Universit\`{a} di Napoli ``Federico II'', Napoli, 80126, Italy}
% ...
\affiliation[6]{Institute of Physics, EPFL, Lausanne, 1015, Switzerland}
% ...
\affiliation[9]{European Organization for Nuclear Research (CERN), Geneva, 1211, Switzerland}
% ...
\affiliation[28]{Millennium Institute for Subatomic physics at high energy frontier-SAPHIR, Santiago, 7591538, Chile}
% ...
\affiliation[33]{Center for Theoretical and Experimental Particle Physics, Facultad de Ciencias Exactas, Universidad Andr\`es Bello, Fernandez Concha 700, Santiago, Chile}
% ...
\affiliation[39]{Department of Physics, Faculty of Science, Chulalongkorn University, Bangkok, 10330, Thailand}
% ...
\affiliation[43]{Affiliated with Pakistan Institute of Nuclear Science and Technology (PINSTECH), Nilore, 45650, Islamabad, Pakistan}
% ...

\emailAdd{guilherme.machado.santos.soares@cern.ch}

\dedicated{© 2026 CERN for the benefit of the \SND{} Collaboration. Reproduction of this article or parts of it is allowed as specified in the CC-BY-4.0 license.}

\abstract{
    We report a measurement of the muon neutrino charged-current (CC)
interaction cross section on tungsten using the electronic detectors of the \SND experiment at the CERN Large Hadron Collider.
The analysis uses proton--proton collision data at a centre-of-mass energy of $\sqrts = 13.6\TeV$, corresponding to an integrated luminosity of
$68.6\fbinv$ collected during LHC Run~3 in 2022 and 2023.
A total of 31 $\numu$CC candidates are selected against an expected
background of $5.0 \pm 1.1$ events, consistent with a signal expectation of $24^{+10}_{-9}$ events. The signal strength is measured to be $\muhat = 1.09^{+0.72}_{-0.37}$, and the combined muon neutrino and anti-neutrino CC cross section on tungsten is determined to be $\sigma(\numu + \numubar) = (37^{+24}_{-12})\times 10^{-35}~\textrm{cm}^2$ at a median energy of $228\GeV$.
In addition, a calorimetric measurement of the hadronic energies of the neutrino candidate events is performed, making use of calibration data from dedicated test-beam campaigns.
}

\begin{document}
\maketitle
\flushbottom

\section{Introduction}
\label{intro}
Recently the Large Hadron Collider (LHC) has been exploited as a source of very high energy neutrinos by means of two dedicated experiments, \SND{}~\cite{SNDLHC:2022ihg} and FASER($\nu$)~\cite{FASER:2022hcn} installed close to the line of sight of the proton beam at a distance of several hundred metres from the proton--proton ($pp$) interaction point. The advent of measurements made with these experiments has led to the `dawn of neutrino physics at colliders'~\cite{Worcester:2023njy} allowing for neutrinos of all flavours to be
detected with high energies.

The \SND experiment consists of a detector placed in the TI18 tunnel at a distance of 480\,m, in the forward region, from the ATLAS interaction point~1 (IP1) of the CERN LHC. The detector is designed to detect and measure interactions of neutrinos produced in decays of particles produced in $pp$ collisions at IP1. The neutrinos produced in the 
forward direction stem from high-energy hadrons and have energies of a few hundred GeV to several~\TeV~\cite{Beni:2019gxv,Beni:2020yfy,Kling:2021gos}, leading to deep inelastic scattering (DIS) in the experiment.

Data taking at the LHC started at the beginning of Run~3 in 2022, and a direct observation of muon neutrino charged-current (CC) interactions  was reported~\cite{SNDLHC:2023pun} based on a data sample of 36.8\,fb$^{-1}$. The FASER experiment reported a similar observation~\cite{FASER:2023zcr}, followed by measurements of the muon neutrino flux and cross section using electronic detectors~\cite{FASER:2024ref},and of the electron and muon neutrino cross sections using an emulsion detector~\cite{FASER:2024hoe}. \SND followed up with a measurement on neutrino interactions with no muons in the final state, thus sensitive to CC interactions of either $\nu_{e}$ or $\nu_{\tau}$ flavour, and neutral-current (NC) neutrino interactions~\cite{SNDLHC:2024qqb}.

This paper presents an updated measurement of $\numu$ CC interactions
using the complete 2022 and 2023 \SND dataset, corresponding to an integrated
luminosity of $68.6\fbinv$, roughly double that of the previous
publication~\cite{SNDLHC:2023pun}. With respect to Ref.~\cite{SNDLHC:2023pun}, the analysis enlarges the fiducial volume (increasing the acceptance of $\numu$ CC interactions in the target from 8\% to 19\%) and introduces a calibrated measurement of the hadronic energy of the
candidate events, enabled by dedicated test-beam campaigns in 2023
and 2024. These improvements allow for the extraction of the combined $\numu$ and $\numubar$ CC cross section averaged over the energy spectrum at the detector, with a median neutrino energy of 228\GeV{}.

%-------------------------------------------------------------------
\section{The SND@LHC detector}
\label{sec:detector}
%-------------------------------------------------------------------
 
\SND{} (Scattering and Neutrino Detector at the LHC) is a hybrid detector combining nuclear emulsion and electronic
detectors~\cite{SNDLHC:2022ihg}.
The present analysis relies exclusively on the electronic subsystems,
which comprise, from upstream to downstream:\\
\textbf{Veto system:} two planes with 7 horizontal plastic
    scintillator bars ($42\times6\times1~\text{cm}^3$), used to tag
    charged particles entering the detector from the IP1 direction.\\
\textbf{Target and SciFi tracker:} five walls, each consisting
    of four emulsion bricks followed by a scintillating-fibre (SciFi) station
    with horizontal and vertical $39\times39~\text{cm}^2$ planes of
    $250~\upmu\text{m}$ polystyrene fibres. During the data-taking period analysed in this paper, the emulsion-instrumented walls contained a luminosity-weighted average of 792~kg of tungsten, which acts as the neutrino interaction target.
    The single-plane spatial resolution of the SciFi detector is approximately $100~\upmu\text{m}$ and the time resolution is $\sim250$\,ps.\\
\textbf{Muon system and hadron calorimeter:} eight scintillator stations interleaved
    with 20\,cm iron absorbers.  The upstream five stations (US) are instrumented with 10 horizontal bars ($85.2\times6\times1~\text{cm}^3$) and act primarily as a hadronic calorimeter. The three downstream stations
    (DS) are equiped with 60 horizontal ($85.2\times1\times1~\text{cm}^3$) and 60 vertical ($60\times1\times1~\text{cm}^3$) scintillator bars to provide muon identification and tracking with a position resolution below 1\,cm and a time resolution of $\sim150$\,ps. The most downstream DS station includes an additional plane of vertical bars.\\

The \SND{} data is collected in trigger-less mode, with all signals exceeding pre-set thresholds read out in 25\,ns time windows. The recorded hits include information on the time of the hit and the integrated digitised charge (QDC).
An online software noise filter reduces the event rate by five orders of
magnitude to $\sim4$\,Hz in the absence of beam, while the detector
operates at $\sim5.4$\,kHz at peak Run~3 luminosity ($2.5\times10^{34}~\text{cm}^{-2}\text{s}^{-1}$). The noise filter induces a small loss in signal efficiency, of around 2\%. In general, each \SND data-taking run corresponds to an LHC fill. In rare occasions, LHC fills are split into several runs.
The detector coordinate system is right handed with its origin at the nominal $pp$ line of sight 480\,m from IP1 in the direction of \SND{}, the $z$ axis pointing away from IP1 towards the \SND{} detector and $y$ pointing against gravity. In this frame, the target region spans [8.0, 57.0]~cm in $x$, [15.5, 54.5]~cm in $y$ and [289, 347]~cm in $z$. The detector is located slightly off the nominal $pp$ collision line of sight, covering a pseudorapidity range of $7.2 < \eta < 8.4$.
  
%-------------------------------------------------------------------
\section{Data and simulation}
\label{sec:data}
%-------------------------------------------------------------------
This analysis makes use of $pp$ collision data collected at the LHC, as well as data from two test-beam campaigns in beamlines derived from CERN's Super Proton Synchrotron (SPS), which are used to calibrate the measurement of the hadronic energy of neutrino interaction candidate events. The detectors exposed to the test beam are reduced, but functionally identical, versions of the \SND{} detector. Monte Carlo (MC) simulations of the LHC accelerator and the \SND{} detector are used to compare the predictions of $pp$ and neutrino interaction event generators to the observed data. 

The \SND readout electronics induce correlated noise which results in delayed hits that are not present in the simulated data. Out-of-time SciFi hits are suppressed by requiring that hits fall within the interval $[-0.5,\,+1.2]$ DAQ clock cycles ($\approx[-3.1,\,+7.5]$\,ns) around a station-wise reference time $t_\mathrm{ref}$. A peak finding algorithm is applied to the SciFi hit time distribution to identify peaks that define $t_\mathrm{ref}$ for each event. Events with bimodal hit time distributions have their $t_\mathrm{ref}$ set to the earliest peak in order to avoid misassignment due to large showers.

\subsection{Proton--proton collision data}
\label{subsec:pp}
 
The dataset consists of all $pp$ collision runs taken at $\sqrts=13.6\TeV$ during stable beam operation in 2022 and 2023, subject to a minimum run duration of 10 minutes and a minimum recorded luminosity of $1~\text{pb}^{-1}$, to ensure the data is collected during periods of stable DAQ and accelerator operation.
A total of 273 runs pass these criteria, corresponding to approximately 91 days of data taking.
The integrated luminosity recorded by the \SND{} electronic detectors was $36.8\fbinv$ (2022) and $31.8\fbinv$ (2023), giving a combined $68.6\fbinv$, using luminosity measurements from the ATLAS Collaboration~\cite{ATLAS:2025kkm,ATLAS:2022hro}, which corresponds to 97.7\% of the delivered luminosity.
 
\subsection{Test-beam data}
\label{subsec:testbeam}
 
Hadronic energy calibration data were collected in two SPS test-beam
campaigns. The 2023 campaign used a replica of the \SND{} US hadronic calorimeter together with an iron mock target instrumented with downsized SciFi modules, subjected to pion beams of 100--300\,GeV~\cite{SNDLHC:2025rph}.
The 2024 campaign deployed a tungsten mock target with half the nominal radiation length in each wall, designed to study electromagnetic shower development and SciFi response. This detector was exposed to 50-300\,GeV electron beams and 180\,GeV pion beams.
 
\subsection{Monte Carlo simulation}
\label{subsec:MC}
 
Neutrino signal events are generated starting from $pp$ collisions
modelled with \textsc{DPMJET-III}~\cite{DPMJET,Roesler:2000he} embedded in a detailed \textsc{FLUKA}~\cite{FLUKA} simulation of the LHC.
Neutrino kinematics extracted at a scoring plane upstream of the detector
are passed to \textsc{Genie} 3.2.0~\cite{Andreopoulos:2009rq,Andreopoulos:2015wxa}, which
models the neutrino interactions and places the interaction vertex in the \SND detector model using a detailed description of its geometry and materials.
Resulting particles are propagated through a \textsc{Geant4}~\cite{Allison:2016lfl} simulation of \SND including
bespoke digitisation models and the online noise filter logic.
The simulated sample corresponds to an integrated luminosity of
$40\abinv$, and a subset of this sample corresponding to $5\abinv$ is used for the systematic uncertainty studies described in Section~\ref{sec:syst} below.
 
Additional predictions are evaluated using a fast
parametrised simulation of the forward neutrino fluxes at the LHC (FASTSIM)~\cite{Kling:2021gos} using three $pp$ generators: \textsc{DPMJET-III}, \textsc{EPOS-LHC}~\cite{Pierog_2015}, and \textsc{SIBYLL}~\cite{Ahn_2009,riehn2015newversioneventgenerator}. The charm-quark contribution is modelled with \textsc{POWHEG}~\cite{Nason:2004rx,Frixione:2007vw,Alioli:2010xd} interfaced with the \textsc{Pythia8} parton shower MC~\cite{Bierlich:2022pfr} in all cases.

Good agreement between \textsc{FASTSIM}/\textsc{DPMJET-III} and the full \textsc{FLUKA}/\textsc{DPMJET-III} simulation is observed in both the neutrino flux and detector-hit multiplicities, validating the use of \textsc{FASTSIM} for the neutrino flux prediction with generators other than \textsc{DPMJET-III}.
 
To overcome the computational challenge of simulating muon-induced neutral-hadron backgrounds, which only very rarely produce neutrino-like signals in the detector, a factorised simulation approach is used to conservatively estimate this contribution. Muon DIS interactions in the materials surrounding the detector (primarily the tunnel walls and floor) are generated with \textsc{Pythia} and propagated with \textsc{Geant4}. The resulting events are filtered to select those that produce signals in the SciFi, while not generating hits in the Veto system. Most of the hits in the SciFi are found to originate from neutrons and neutral kaons, but none of these events survive the full set of criteria used to select neutrino interaction candidates. Therefore, a second simulation step is performed to conservatively estimate the efficiency of selecting neutron and neutral kaon events as neutrino interaction candidates. Neutron and neutral kaon particle gun events are generated with \textsc{Geant4}, with the particles uniformly distributed across the upstream face of the detector and their energies distributed according to the energy spectra of neutrons and neutral kaons reaching the target in the full muon DIS simulation. With around 10$^{7}$ neutral hadron particle gun events generated, the probability for selecting this background into the signal sample is estimated to be around 10$^{-6}$. However, given the low rate of entering neutral hadrons expected from the muon DIS simulations, this forms a subdominant contribution to the total background, as described in Section~\ref{sec:neutral_neutrino_background}.
 
%-------------------------------------------------------------------
\section{Event selection}
\label{sec:selection}
%-------------------------------------------------------------------
 
Signal candidates are separated from the large muon background
using a cut-based procedure.
The overall data reduction factor is $5.5\times10^{8}$, yielding an
 efficiency for selecting $\numu$ CC interactions in the target of 7\%. The number of data events passing each cut is given in Table~\ref{tab:cutflow} together with the signal efficiency obtained from the MC. A detailed description of each selection cut is provided below, separately for data quality cuts that do not apply to the MC, fiducial volume acceptance, and the $\numu$ CC signal identification criteria.
 
\begin{table}[htbp]

\centering
\begin{tabular}{clcc}
\toprule
Cut & Description                          & Data events & $\numu$CC \\
    &                                      &             & eff. \\
\midrule                     
 & All selected runs                       & $1.69\times10^{10}$ & --- \\
 & Stable beams                            & $1.66\times10^{10}$ & --- \\
 & IP1 bunch crossing                      & $1.54\times10^{10}$ & --- \\
 & Time interval to prev. event            & $1.14\times10^{10}$ & --- \\
A & SciFi fiducial area                    & $8.44\times10^{8}$ & 0.43 \\
B & DS fiducial area                       & $7.23\times10^{8}$ & 0.81 \\
C & No Veto hits                           & $1.31\times10^{6}$ & 0.62 \\
D & No hits in first SciFi                 & $8.05\times10^{5}$ & 0.86 \\
\midrule
E & $>35$ SciFi hits                       & $7.79\times10^{4}$ & 0.86 \\
F & $Q_{\mathrm{US}}>600$ units                    & $1.86\times10^{4}$ & 0.94 \\
G & All US planes hit                      & 8592 & 1.00 \\
H & $N_{\textrm{DS}}^{\textrm{eff}} \leq 10$               & 4781 & 0.71 \\
I & $\geq1$ DS track                  & 853 & 0.96 \\
J & IP1 timing consistency                 & 840 & 1.00 \\
K & DS track in fiducial volume           & 59 & 0.88 \\
L & Avg.\ DOCA $<3$\,cm                    & \textbf{31} & 0.74 \\
\midrule
 & Fiducial volume acceptance (cuts A--D) & & 0.19 \\
 & Selection efficiency (cuts A--L) & & 0.07\\
\bottomrule
\end{tabular}
\caption{Analysis event selection cut flow, with associated event yields in
data. The cut efficiencies, determined from $\numu$ CC signal MC, are given relative to the
preceding cut.
}
\label{tab:cutflow}
\end{table}
 
\paragraph{Data quality.}
Events in the runs selected for analysis according to the criteria defined in Section~\ref{subsec:pp} are further required to fulfill the following data quality criteria: the event must occur during the stable beams stage of the LHC fill, be consistent with two proton bunches crossing at IP1 within the 6.25~ns DAQ clock granularity, and have a time interval of at least 100 DAQ clock cycles (625~ns) to the previous event. The last criterion is applied to circumvent the delayed noise observed in the data, which can lead to spurious noise-only events following a physical event. This cut has a negligible impact in the signal efficiency, of around 0.1\%.

\paragraph{Fiducial acceptance (cuts A--D).}
A $25\times26~\text{cm}^2$ fiducial cross-sectional area is defined using average SciFi hit channel (cut~A) and DS hit bar number (cut~B) ranges.
Events are further required to contain no Veto hits (cut~C) and no hits in the first SciFi station, counted from upstream to downstream, (cut~D), selecting interactions consistent with a neutral particle interaction in walls 2--5 of the target. This is the main change in the event selection criteria with respect to Ref.~\cite{SNDLHC:2023pun}, where only walls 3 and 4 were included in the fiducial volume in order to achieve a lower background.
The combined efficiency of the fiducial volume cuts is 19\%, given by the ratio between the number of $\numu$ CC events passing cuts A--D and all $\numu$ CC interactions simulated in the tungsten target.
 
\paragraph{Signal identification (cuts E--L).}
The identification of $\numu$~CC interaction candidates relies on the requirement that the activity in the SciFi and hadron calorimeter is consistent with a hadron shower and significantly larger than what is expected from muon-induced backgrounds, combined with the reconstruction of a track in the DS system, which indicates the presence of an outgoing muon. Additional cuts are used to ensure the quality of the reconstructed track, as well as the consistency between the track and the hadron shower.

The hadron shower criteria consist of the requirement of more than 35 SciFi hits (cut~E) and a total US QDC ($Q_{\mathrm{US}}$) larger than 600 units (cut~F). It is further required that all five US planes contain hits, as is expected from an event with a muon originating in the target (cut~G).

Events are required to have a maximum of 10 effective DS hits, $N_{\textrm{DS}}^{\textrm{eff}} \leq 10$, with
\[
N_{\textrm{DS}}^{\textrm{eff}} = \left(\sum_{i = 1}^3 \frac{n_i^{\textrm{ver}} + n_i^{\textrm{hor}}}{2}\right) + n_4^{\textrm{ver}}~,
\]
\noindent where $n_i^{\textrm{ver}}$ and $n_i^{\textrm{hor}}$ are the number of hits in the vertical and horizontal bars of the $i^{\textrm{th}}$ plane, respectively. The last term, $n_4^{\textrm{ver}}$, is the number of hits in the 4$^\textrm{th}$ vertical plane, which does not have a matching horizontal plane. This cut (H) is used to discard events where the high multiplicity of hits in the DS can lead to track reconstruction ambiguity.

At least one reconstructed DS track must be identified (cut~I) using a Hough transform algorithm for pattern matching and a Kalman filter tracking procedure~\cite{SNDLHC:2023mib}. A minimum of 3 planes hit in each projection is required for track candidates to be formed.

Events are required to originate from the IP1 direction (cut~J) by requiring that the time difference between the latest DS hit and the earliest SciFi hit is positive.

Finally, it is required that the DS track intercepts the analysis fiducial area when extrapolated to the Veto detector (cut~K) and that the distance of closest approach (DOCA) of the extrapolated DS track to the hits in the SciFi is smaller than $3$\,cm when averaged over each station (cut~L).

\begin{figure}
  \centering
  \includegraphics[width=1.\textwidth]{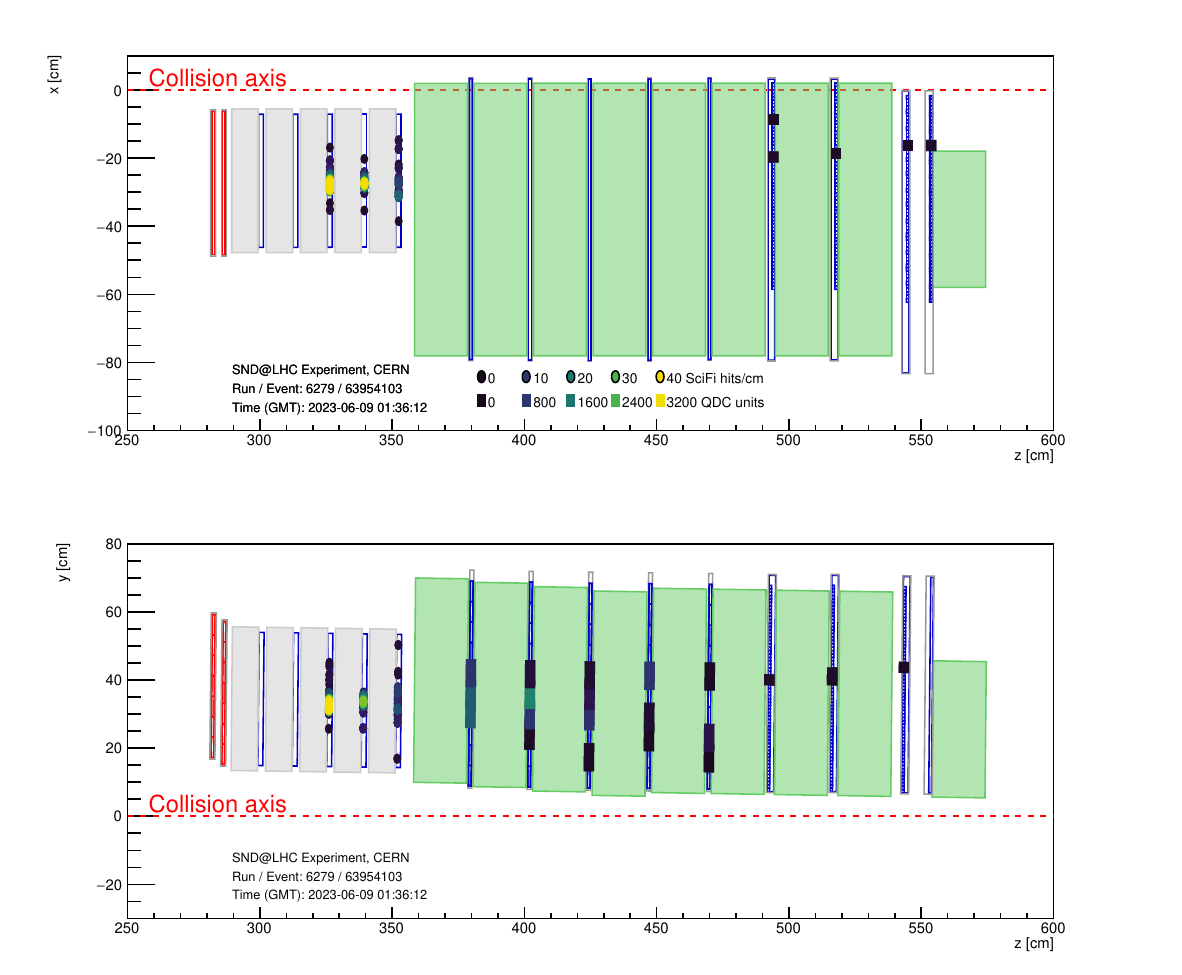}
\caption{A $\numu$ CC interaction event candidate in the 2023 data. Top-down (top) and side (bottom) views of the detector are shown, with IP1 particles arriving from the left side of the figure. From left to right, the Veto detector is shown in red, the tungsten-emulsion target walls in grey and the US hadron calorimeter and DS muon system iron absorbers in green. SciFi, US, and DS sensitive planes are shown in blue, with hits in these detectors represented by markers coloured according to the hit density~\cite{SNDLHC:2024qqb} (SciFi) or QDC value (US and DS). The candidate neutrino interaction occurs roughly in the middle of the third target wall with high SciFi and US activity indicating the presence of a large hadron shower, and a clear track in the DS indicating the outgoing muon. The event has a reconstructed hadronic energy of 0.24$\pm$0.03\TeV{}.}
\label{fig:event_display}       % Give a unique label
\end{figure}

After the full selection, 31 $\numu$ CC candidate events are identified: 13 from 2022 and 18 from 2023. All events included in the analysis of 2022 data reported in Ref.~\cite{SNDLHC:2023pun} are included in this set, except for one event which migrated out of the sample after an update to the tracking algorithm. An event display of one of the $\nu_\mu$ CC interaction candidates recorded in 2023 is shown in Figure~\ref{fig:event_display}.
 
%-------------------------------------------------------------------
\section{Background estimation}
\label{sec:background}
%-------------------------------------------------------------------
The background is dominated by two classes of events: muons
that enter the detector from the sides without traversing the Veto
(side-entering muons), and muons that traverse the detector without
activating the Veto or first SciFi plane due to detector inefficiency
(penetrating muons). Both are estimated with data-driven methods.

Additionally, muon DIS interactions in the material surrounding the detector can generate neutral hadrons that interact or decay within the target. These hadrons have typically significantly lower energy than neutrinos~\cite{SNDLHC:2023pun} and rarely produce muons in the analysis acceptance. This subdominant background contribution is conservatively estimated with the simulation-based methods described in Section~\ref{subsec:MC}.
Finally, neutrino NC interactions and CC interactions of flavours different from $\numu$ and $\numubar$ result in a small background estimated from MC.

Details of each of these background contributions are given below.

\subsection{Side-entering muon background}

The side-entering muon background is estimated with the data-driven ``ABCD" method. Three sideband control regions (A, B and C) are defined by inverting cut~F ($Q_{\mathrm{US}}$ threshold) and cut~K (DS track in fiducial volume) as described in Table~\ref{tab:abcd}, with the signal region~D defined by the standard cuts.

\begin{table}[htbp]
  \centering
\begin{tabular}{ c c c c c }
\toprule
Region & $Q_{\mathrm{US}}$ (cut F) & DS track in FV (cut K) & Data & Neutrino MC \\ \midrule
Control \textbf{A} & $<$ 600 units& True  & 12 & 2.14 \\
Control \textbf{B} & $<$ 600 units& False & 564 & 0.2 \\
Control \textbf{C} & $>$ 600 units& False & 185 & 2.26 \\
Signal \textbf{D} &  $>$ 600 units& True & 31 & 19.9 \\
\bottomrule
\end{tabular}
\caption{Definition of control regions with the $Q_{\mathrm{US}}$ and DS track in the fiducial volume (FV) used to estimate the side-entering muon background, including the number of data events in each of the regions and the number of neutrino events predicted by \textsc{DPMJET-III} using the complete \textsc{FLUKA} LHC simulation.}
\label{tab:abcd}
\end{table}

Under the assumption that the US energy deposition is independent of the DS track direction for muon events, the background in the signal region is:
\begin{equation*}  
N_\mu^{\mathrm{side}} =
    \frac{(N_A^{\mathrm{data}} - N_A^{\nu\mathrm{MC}})
          (N_C^{\mathrm{data}} - N_C^{\nu\mathrm{MC}})}
         {N_B^{\mathrm{data}} - N_B^{\nu\mathrm{MC}}}~,
\end{equation*}
where the expected neutrino contribution in control region $i$, $N_i^{\nu\mathrm{MC}}$, is subtracted from the observed data in the same region, $N_i^{\mathrm{data}}$. For this background estimate, the neutrino flux predicted by \textsc{DPMJET-III} and the full \textsc{FLUKA} simulation of the LHC are used. 

\begin{figure}
  \centering
  \includegraphics[width=1.\textwidth]{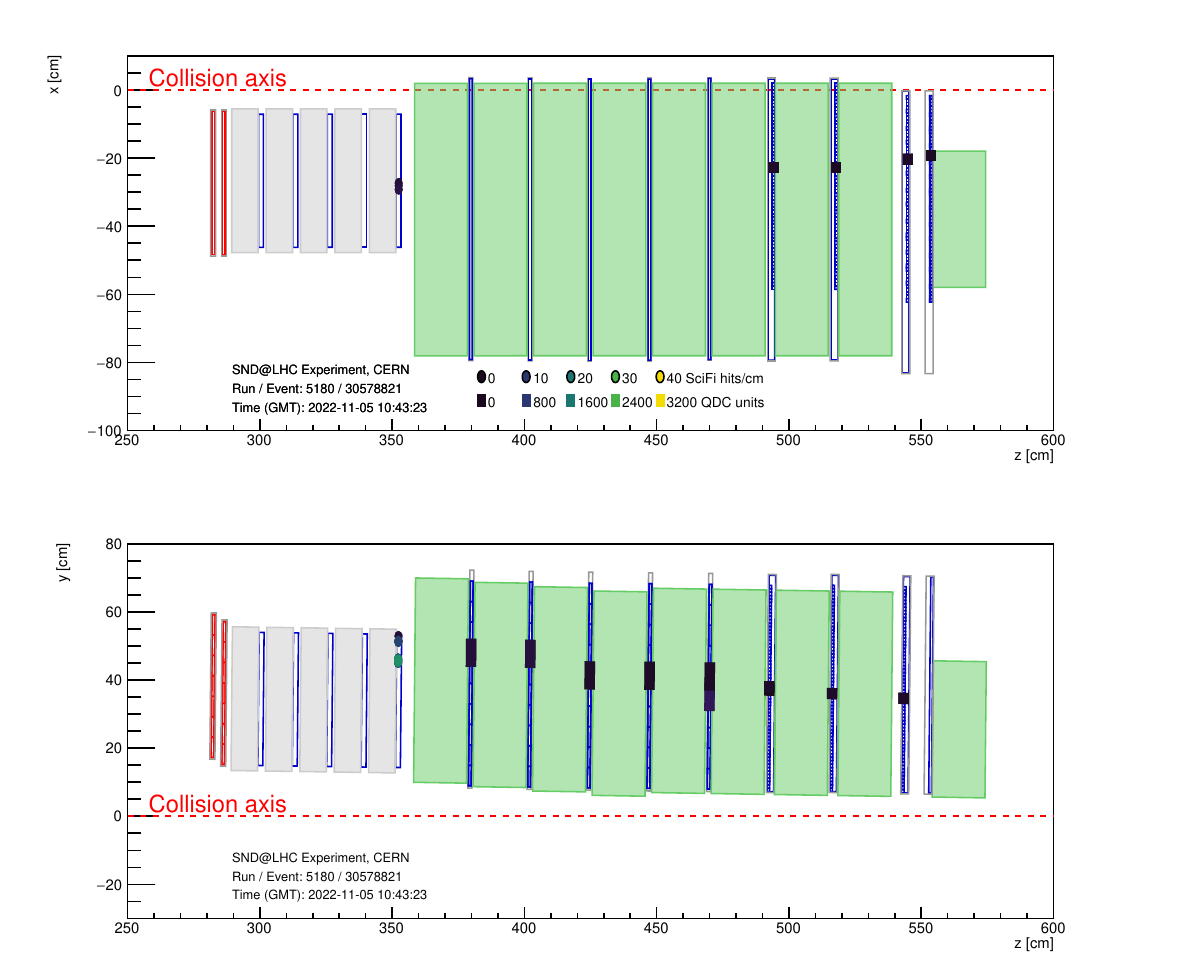}
\caption{A selected neutrino candidate event that appears to be a side-entering muon on visual inspection. Hits very close to the edge of the detector in the fifth SciFi plane are a possible indication of a charged particle entering the \SND{} target at this point and therefore not traversing the Veto and first SciFi plane.}
\label{fig:event_display_sidemu}       % Give a unique label
\end{figure}

The uncertainty in the predicted number of side-entering muon events in the signal region is dominated by the small number of data events in region~A, which leads to an uncertainty of 34\%. The resulting estimate is $N_\mu^{\mathrm{side}} = 3.2 \pm 1.1$\,events. The uncertainty on this background is larger than the difference between the prediction of \textsc{DPMJET-III} and other $pp$ collision generators, and therefore this choice of event generator for the calculation of $N_\mu^{\mathrm{side}}$ is robust. An event selected into the signal sample that on visual inspection appears to be a side-entering muon is shown in Figure~\ref{fig:event_display_sidemu}.

\subsection{Penetrating muon background}
Penetrating muons traversing the Veto and first SciFi plane can induce backgrounds due to the small but finite inefficiency of the detectors. This background component is estimated by applying a reduced version of the event selection cuts, with the Veto and first SciFi plane cuts removed, and combining the result with the known muon flux and detector inefficiencies. The penetrating muon background is given by:
\begin{equation*}
  N_\mu^{\mathrm{pen}} = N_\mu \cdot \eta_\mu \cdot
    \varepsilon_V \cdot \varepsilon_{\mathrm{SF}}~,
\end{equation*}
where $N_\mu$ is the muon yield, $\eta_\mu$ is the survival probability of the muon sample through the event selection with cuts~C and~D omitted,
and $\varepsilon_V$, $\varepsilon_{\mathrm{SF}}$ are the Veto
and first SciFi plane inefficiencies, respectively.

The total number of events in the data is used as a proxy for the muon yield $N_\mu$, corresponding roughly to the muon flux at the experiment integrated over the analysed runs. The measurement of the muon flux at \SND is reported in Refs.~\cite{SNDLHC:2023mib,SNDLHC:2026mip}, remaining at a constant level of $\left(2.1\pm0.1\right)\times10^{-2}$~nb/cm$^2$ in 2022 and 2023.

Applying the selection of Table~\ref{tab:cutflow} except cuts C and D yields a passing muon survival probability $\eta_\mu = [0.4, 6.4]\times10^{-3}$, with the range representing the variation due to changes in the DAQ system configuration throughout the data taking period.

The Veto inefficiency varied significantly across data-taking periods~\cite{Ruf:2860187,Ruf:2880073}.
Until late October 2022, misaligned Veto readout boards caused
inefficiencies of order $10^{-3}$--$10^{-4}$. After realignment and
noise filter improvements the inefficiency dropped to $10^{-7}$.
The first SciFi plane inefficiency remained stable throughout the analysed period, with a measured inefficiency between $1.1\times10^{-4}$ and $2.5\times10^{-4}$~\cite{Ruf:2872240,Ruf:2875738}. The much lower inefficiency of the Veto system compared to the SciFi plane stems from the former being a combination of two staggered detector planes with no geometrical dead-space, while the latter consists of a single plane with gaps between the sensitive elements of the detector having an impact on its efficiency.

To estimate the penetrating muon background, 9 data-taking periods are defined to take into account variations in the detector conditions throughout 2022 and 2023~\cite{teseSoares}. The penetrating muon rate is estimated separately for each period, with the period corresponding to the high Veto system inefficiency having the largest contribution to the background yield. The total background due to penetrating muons is estimated to be $N_\mu^{\mathrm{pen}} = 1.37 \pm 0.23$\,events, with the uncertainty derived from the variation of the muon survival probability throughout each data-taking run.

\subsection{Neutral hadron and neutrino backgrounds}
\label{sec:neutral_neutrino_background}
The neutral hadron background is estimated with a simulation of muon DIS interactions in the material surrounding the experiment, most importantly the tunnel floor and walls, corresponding to around 20$\fbinv$. Muon DIS interactions where particles produce signals in the \SND target in events without Veto hits are analysed to extract the spectra of neutral hadrons interacting in the target. It is found that these interactions are dominated by neutrons and neutral kaons with energies up to 100\GeV{}. Particle gun simulations of events with these characteristics are used to estimate the efficiency for selecting neutral hadron interactions as signal candidates. The estimated background due to neutral hadrons is $0.3 \pm 0.1$ events, and therefore much smaller than the data-driven estimates of the muon backgrounds described above.

Finally, neutrino interactions of non-muon flavours are found to contribute 0.1 events to the background, according to the \textsc{DPMJET-III} prediction and full \textsc{FLUKA} simulation of the LHC.

The total expected background is $5.0 \pm 1.1$ events, with the breakdown summarised in Table~\ref{tab:backgrounds}.
\begin{table}[htbp]

\centering
\begin{tabular}{lc}
\toprule
Background component & Events \\
\midrule
%$\numu$ CC signal (EPOS-LHC + POWHEG) & $24^{+10}_{-9}$ \\
Side-entering muons & $3.2 \pm 1.1$ \\
Penetrating muons & $1.4 \pm 0.2$ \\
Neutral hadrons & $0.3 \pm 0.1$ \\
Other neutrino flavours & $0.1$ \\
\midrule
Total & $5.0\pm1.1$ \\
\bottomrule
\end{tabular}
\caption{Summary of expected event yields per background component.}
\label{tab:backgrounds}
\end{table}
%-------------------------------------------------------------------
\section{Systematic uncertainty on the signal efficiency}
\label{sec:syst}
%-------------------------------------------------------------------
 
Systematic uncertainties on the signal efficiency are evaluated by varying each selection criterion by an amount informed by data--MC comparisons.

To estimate the uncertainty from the SciFi fiducial area cut (A), test-beam events with pion-induced hadronic showers are used to compare the reconstructed position between data and MC. A difference of around 0.5~cm in the resolution of the shower axis is observed, which translates into an uncertainty from the SciFi fiducial area of [-7.1, 5.4]\%. For the DS fiducial area cut (B), muon events in the $pp$ data are reconstructed in the SciFi to estimate the DS position reconstruction resolution in both data and MC. A difference of 0.6~cm is observed in the distance between DS hit positions and the extrapolated SciFi track, leading to an uncertainty of [-3.8, 2.6]\%.

The uncertainty from the simulation of the event builder logic is estimated by varying the nominal 25~ns event window by [-0.4,+1.0]~ns to reflect differences in the hit time distributions between neutrino candidate events in the data and MC. Slow particles produced in neutrino interactions are observed in the MC to produce hits in the Veto and first SciFi stations, upstream of the interaction vertex. Changes in the time window can lead to these Veto and first SciFi hits migrating out of events, which in turn leads to changes in the signal efficiency for cuts C and D. The impact of event window variations on the signal efficiency estimate is [-0.3, 0.8]\%.

Test-beam data is used to estimate the uncertainty from the minimum activity cuts in the SciFi (E) and US hadron calorimeter (F). The uncertainty is estimated by applying the minimum $Q_{\mathrm{US}}$ cut to measure data--MC differences in the number of SciFi hits, and then reversing the procedure by applying the required minimum number of SciFi hits and analysing the $Q_{\mathrm{US}}$ distribution. A difference of $\pm4$ SciFi hits is observed in the test-beam pion events, leading to a signal efficiency uncertainty of $\pm2.3$\%. Known shortcomings of the $Q_{\mathrm{US}}$ digitisation model in the MC lead to large correction factors being derived from the comparison of the distribution of $Q_{\mathrm{US}}$ values for test-beam pion events. The largest correction factor, of around 4, is observed for 300\GeV{} pions (the highest energy available in the test-beam data) and leads to a one-sided systematic uncertainty of [-33, 0]\%.

The uncertainty from $N_{\textrm{DS}}^{\textrm{eff}}$ is estimated by comparing the distributions of the number of DS hits in 300\GeV{} data and MC test-beam pion events. A systematic difference of 1 hit is observed between the data and MC, resulting in an uncertainty of [-4.1, 2.6]\%.

Track reconstruction efficiencies are found to be systematically overestimated in the MC by 3.7\%~\cite{SNDLHC:2023mib}. This translates to a one-sided uncertainty of [-3.7, 0]\%. Passing muon events in the $pp$ collision data and MC are used to estimate the uncertainty on the track extrapolation quality cuts K and L. To estimate the uncertainty associated to the extrapolation of the DS track to the SciFi fiducial area (K) the distribution of the distance between the extrapolation of the DS track to the Veto plane and the closest Veto hits is compared between data and MC. A difference of 0.1~cm is obtained resulting in a systematic uncertainty of [-0.15, +0.5]\%. Cut L relies on the DOCA of the extrapolated DS track to hits in the SciFi and therefore this cut is sensitive to the presence of showers in the SciFi planes. To study the systematic uncertainty from this cut, a modified event selection is applied to the $pp$ collision data, where the veto cuts (C and D) are omitted as well as the DOCA cut (L). This yields a high-statistics sample of events with sizeable activity in the SciFi. A difference of 0.3~cm in the DOCA is observed between data and MC, resulting in a systematic uncertainty of [-5.4, 5.0]\%.

The systematic uncertainties on the signal efficiency are summarised in Table~\ref{tab:syst}. The total uncertainty is [-34.9, 8.6]\% with the negative side dominated by the mismodelling of the $Q_{\mathrm{US}}$ distribution and the positive side dominated by uncertainties on the SciFi hit distribution.
 
\begin{table}[htbp]
\centering
\begin{tabular}{lc}
\toprule
Source & Uncertainty (\%) \\
\midrule
SciFi fiducial area            & [-7.1, +5.4] \\
DS fiducial area               & [-3.8, +2.6] \\
Event timing window            & [-0.3, +0.8] \\
Minimum SciFi hits             & $\pm2.3$ \\
Minimum $Q_{\mathrm{US}}$      & [-33.0, +0] \\
Maximum DS hits                & [-4.1, +2.6] \\
Presence of DS track           & [-3.7, +0] \\
Track extrapolation to SciFi~1 & [-0.15, +0.6] \\
Maximum average DOCA           & [-5.4, +5.0] \\
\midrule                              
Total                          & [-34.9, +8.6]\\
\bottomrule
\end{tabular}
\caption{Systematic uncertainties on the $\numu$CC signal efficiency.}
\label{tab:syst}
\end{table}
 
%-------------------------------------------------------------------
\section{Hadronic energy measurement}
\label{sec:Ehad}
%-------------------------------------------------------------------
  
The hadronic shower energy is reconstructed from a linear combination of
the QDC signals in the SciFi tracker and the US
calorimeter~\cite{SNDLHC:2025rph}:
\begin{equation*}
  E_{\mathrm{tot}} = k\cdot Q_{\mathrm{SciFi}} + \alpha\cdot Q_{\mathrm{US}}~,
  \label{eq:Eline}
\end{equation*}
where $k$ and $\alpha$ are calibration constants.
These are derived separately for each shower-start SciFi station using
principal component analysis of the anti-correlated $Q_{\mathrm{SciFi}}$
vs.\ $Q_{\mathrm{US}}$ distribution, observed for pion beams of 100, 180,
and 300\,GeV in the 2023 test-beam campaign.
 
Since the 2023 test-beam SciFi modules were equipped with reflective mirrors
on the non-readout end, which are absent in \SND and in the 2024 test-beam detector, a correction factor of ${\simeq}2.6$ is applied to $Q_{\mathrm{SciFi}}$, derived by comparing 180\,GeV pion signals between the 2023 and 2024 test-beam
datasets.
  
The achieved hadronic energy resolution ranges from 22\% at 100\,GeV to
12\% at 300\,GeV. 
The calibrated energy formula given above is applied to all 31 $\numu$ CC candidate events. Reconstructed hadronic energies range from a few $\GeV$ (consistent with residual passing-muon contamination) to a maximum of $390\GeV$.
The reconstructed hadronic energy is broadly consistent with the MC expectation from \textsc{EPOS-LHC} and \textsc{POWHEG}, as shown in Figure~\ref{fig:had_energy}.
 
%-------------------------------------------------------------------
\section{Results}
\label{sec:results}
%-------------------------------------------------------------------
 
A total of 31 events pass the signal selection criteria, with a background expectation of 5.0$\pm$1.1 events. The $\numu$ signal expectation is obtained for different $pp$ collision generators by reweighting the \textsc{DPMJET-III} MC events as a function of neutrino energy. 

The expected flux of neutrinos originating in the decay of light hadrons is estimated with the \textsc{SIBYLL}, \textsc{DPMJET-III} and \textsc{EPOS-LHC} generators, which, together with the charm hadron component estimated with \textsc{POWHEG}, give a total expected signal yield of 19.3, 20.4 and 24.2 events, respectively. Following Ref.~\cite{FASER:2024ykc}, we take the expectation from \textsc{EPOS-LHC} as the nominal prediction for the light-hadron component, assuming half of the generator spread as the uncertainty on the signal yield originating in light-hadron decays. The uncertainty on the charm-hadron component is obtained by varying the factorisation and resummation scales in \textsc{POWHEG} according to the procedure in Ref.~\cite{Buonocore:2023kna}. This results in the dominant uncertainty on the signal production, with a charm-hadron component of $5.7^{+9.2}_{-2.5}$ events.

The expected signal yield is 24$^{+10}_{-3}$ events, which together with the background expectation of 5.0$\pm$1.1 events gives a total of 29$^{+10}_{-3} (\textrm{prod.})^{+2}_{-8}(\textrm{eff.})\pm1(\textrm{bkg.)}$ events, with uncertainties associated to the signal production (prod.), selection efficiency (eff.), and background expectation (bkg.), respectively. This expectation is to be compared with 31 events observed in the data.

\begin{figure}[hbtp]
    \begin{center}
    \includegraphics[width=0.5\linewidth]{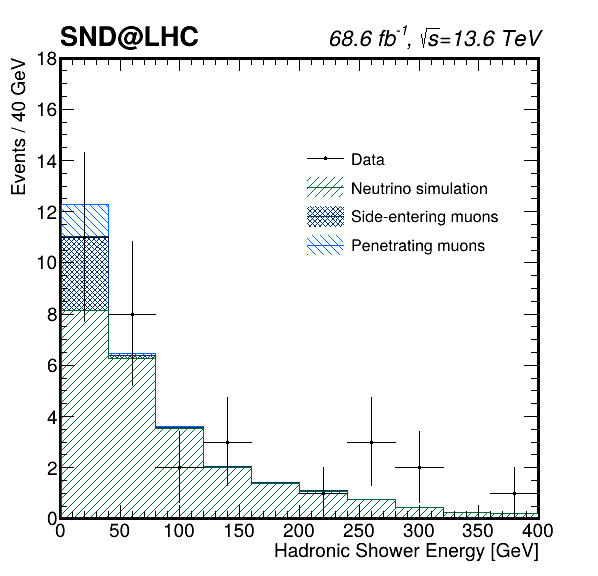}
    \end{center}

    \caption{Measured hadronic shower energy distribution overlaid with model expectations.}
    \label{fig:had_energy}
\end{figure}

\begin{figure}[hbtp]
    \begin{center}
    \includegraphics[width=0.5\linewidth]{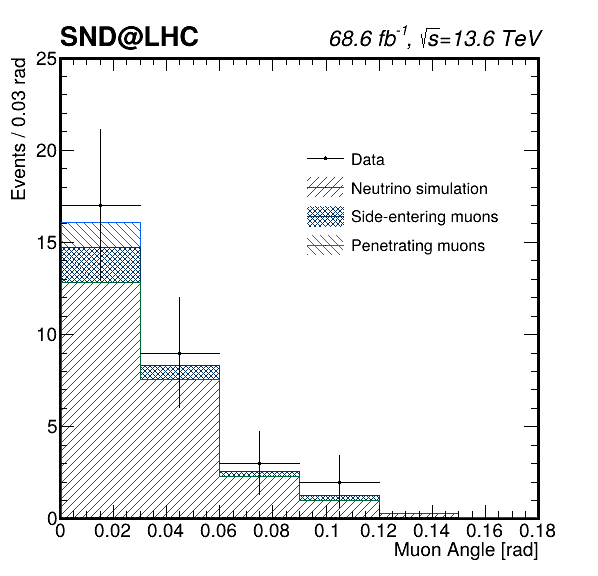}
    \end{center}
    \caption{Angle of the reconstructed muon, overlaid with model expectations.}
    \label{fig:MuonEntryAngle}
\end{figure}

The measured hadronic energy spectrum, featuring deposited
energies up to ~0.4\TeV{}, is presented in  Figure~\ref{fig:had_energy}. This result demonstrates the calorimetric capability of the SND@LHC electronic detector system and provides the first measurement of this observable in collider neutrino interactions. The corresponding muon angle distribution is
shown in Figure~\ref{fig:MuonEntryAngle}, exhibiting good agreement with the expectation.

The number of observed candidates is compared with the expected signal plus background yields using a log-likelihood fit following the formalism of Ref.~\cite{Cowan:2010js} implemented in Refs.~\cite{pyhf,pyhf_joss}.
The signal strength $\mu$ parameterises the ratio of observed to
expected signal yield, with systematic uncertainties incorporated as nuisance parameters constrained by Gaussian probability distributions. The best-fit signal strength is $\muhat = 1.09^{+0.72}_{-0.37}$.

\begin{figure}[hbtp]
    \begin{center}
    \includegraphics[width=0.5\linewidth]{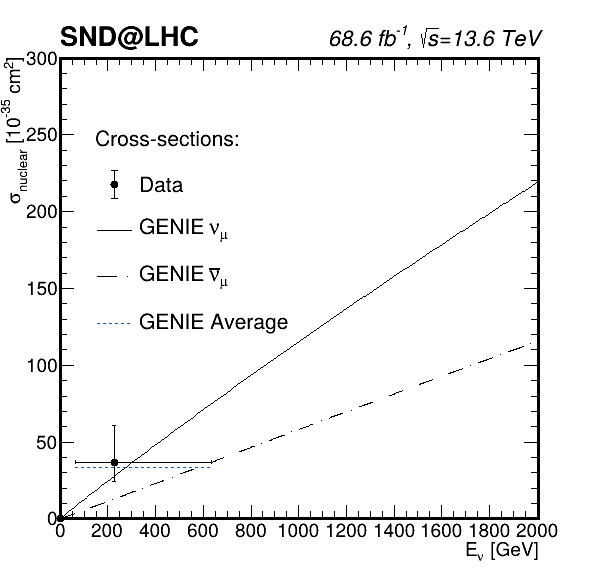}
    \end{center}
    \caption{Measured muon neutrino and anti-neutrino combined cross section for CC interactions on tungsten, and comparison with model predictions. The expected fractions of $\nu_\mu$ and $\bar{\nu}_\mu$ are 68.1\% and 31.9\%, respectively.
    The error bars of the measured cross section represent the $[-1\sigma,+1\sigma]$ interval obtained through the log-likelihood function, while the horizontal bars represent the 15.87 and 84.14 percentiles for the (anti)neutrino energy distribution according to the estimated \textsc{EPOS-LHC} and \textsc{POWHEG} fluxes, with the data point drawn at the median energy. The solid and dashed black lines represent the \textsc{Genie} predicted cross section as a function of neutrino energy for muon neutrinos and antineutrinos, respectively. The dotted blue line shows the \textsc{Genie} cross section averaged over the $\numu$ and $\numubar$ fluxes.}
    \label{fig:xsec}
\end{figure}

The expected cross section for combined $\numu$ and $\numubar$ CC interactions on tungsten from \textsc{Genie} averaged over the flux at \SND{} is $\sigma_\text{Expected} = 34\times 10^{-35}~\text{cm}^2$. The measured cross section at \SND{} is extracted from the signal strength:
%The cross section for combined $\numu$ and $\numubar$ CC interactions on tungsten is extracted from the signal strength and the \textsc{Genie} cross-section model used to generate the events:
\begin{equation*}
  \sigma(\numu + \numubar) =
    (37^{+24}_{-12})\times10^{-35}~\text{cm}^2.
\end{equation*}
The average (anti)neutrino energy according to the \textsc{EPOS-LHC} and \textsc{POWHEG} fluxes is $343\GeV$, with a median of $228\GeV$ and a central 68\% interval of $[64,\,635]\GeV$. The expected fractions of $\nu_\mu$ and $\bar{\nu}_\mu$ in the analysis acceptance are 68.1\% and 31.9\%, respectively.
As shown in Figure~\ref{fig:xsec}, the measured cross section is consistent with the \textsc{Genie} prediction.
 
%-------------------------------------------------------------------
\section{Conclusions}
\label{sec:conclusions}
%-------------------------------------------------------------------
 
We have presented an updated measurement of muon neutrino charged-current interactions, employing proton–proton collision data at a centre-of-mass energy $\sqrt{s}= 13.6$\TeV{} collected by the SND@LHC experiment during LHC Run 3 in 2022 and 2023, corresponding to an integrated luminosity of $68.6\fbinv$. The analysis benefits from a doubling of the integrated luminosity relative to Ref.~\cite{SNDLHC:2023pun}, an expanded fiducial volume
(acceptance increased from 8\% to 19\%).

Thirty-one $\numu$ CC candidates are observed, consistent with a
signal expectation of $24^{+10}_{-9}$ events and a background of
$5.0 \pm 1.1$ events.
The signal strength $\muhat = 1.09^{+0.72}_{-0.37}$ reflects the agreement between observation and theory prediction.
The measured charged-current interaction cross section on tungsten
is $\sigma(\numu + \numubar) = (37^{+24}_{-12})\times10^{-35}~\text{cm}^2$
at a median energy of $228\GeV$.
Additionally, we report the first calorimetric energy measurement of collider neutrino interactions, including events with hadronic energies of up to 0.4\TeV{}, in good agreement with the predictions.
 
Future analyses will benefit from the significantly upgraded Veto
system installed at the start of 2024 data taking~\cite{SNDLHC:2025nrj,tesiPaggi},
continued developments in detector calibration and MC
simulation, and the substantially larger dataset expected from the
remainder of Run~3.
These improvements are expected to allow for a significantly larger fiducial volume leading to better statistical precision as well as a reduction of the dominant systematic uncertainty associated with hadronic shower modelling.
The calorimetric hadronic energy measurement demonstrated in this paper is a precursor for the higher-precision studies of collider neutrino interactions enabled by the fully-electronic silicon microstrip upgrade of the \SND detector~\cite{Abbaneo:2926288,SNDLHC:2026why} planned for the high-luminosity Run~4 of the LHC.

\acknowledgments
We would like to thank Felix Kling for his assistance with producing predictions of different $pp$ collision generators using the \textsc{FASTSIM} tool. We thank Luca Buonocore and Francesco Tramontano for their help with using \textsc{POWHEG} for generating the charm hadron neutrino flux predictions for \SND.

We acknowledge the support for the construction and operation of the SND@LHC detector provided by the following funding agencies:  CERN;  the Bulgarian Ministry of Education and Science within the National Roadmap for Research Infrastructures 2020–2027 (object CERN); ANID FONDECYT grants No. 3230806, No. 1240066, 1240216 and ANID  - Millenium Science Initiative Program - $\rm{ICN}2019\_044$ 
(Chile); the Deutsche Forschungsgemeinschaft (DFG, ID 496466340); the Italian National Institute for Nuclear Physics (INFN); JSPS, MEXT, the~Global COE program of Nagoya University, the~Promotion and Mutual Aid Corporation for Private Schools of Japan for Japan;
the National Research Foundation of Korea with grant numbers
2021R1A2C2011003, 2020R1A2C1099546,
2021R1F1A1061717, and 2022R1A2C100505; Fundação
para a Ciência e a Tecnologia (FCT, Portugal) grant numbers  
CEECIND/01334/2018, CEECINST/00032/2021,\\ PRT/BD/153351/2021
and CERN/FIS-INS/0028/2021; the Swiss National Science Foundation (SNSF); TENMAK for Turkey (Grant No. 2022TENMAK(CERN) A5.H3.F2-1).
J.C.~Helo~Herrera and O.~J.~Soto~Sandoval acknowledge support from ANID 
FONDECYT grants No.1241685 and 1241803.
M.~Climesu, H.~Lacker and R.~Wanke are funded by the Deutsche Forschungsgemeinschaft (DFG, German Research Foundation), Project 496466340. This research was financially supported by the Italian Ministry of University and Research within the Prin 2022 program.

We express our gratitude to our colleagues in the CERN accelerator departments for the excellent performance of the LHC. We thank the technical and administrative staff at CERN and at other \SND institutes for their contributions to the success of the \SND efforts. We thank Luis Lopes, Jakob Paul Schmidt and Maik Daniels for their help during the~construction.

\bibliographystyle{JHEP}
\bibliography{numu2223_jhep}
\end{document}